\documentclass[preprint,showpacs,preprintnumbers,amsmath,amssymb,superscriptaddress]{revtex4}

\usepackage{graphicx}
\usepackage{dcolumn}
\usepackage{bm}

\newcommand{\etal} {{\it et al.} }
\newcommand{\wavno} {cm$^{-1}$}

\begin{document}


\title{The Structure of Rh$_8^+$ in the Gas Phase}

\author{D.~J.~Harding}
\altaffiliation[Current address: ]{Fritz-Haber-Institut der Max-Planck-Gesellschaft, Faradayweg 4-6, D-14195 Berlin, Germany}
\affiliation{Dept. of Chemistry, University of Warwick, Coventry, CV4 7AL, U.K.}
\author{T.~R.~Walsh}
\affiliation{Dept. of Chemistry and Centre for Scientific Computing, University of Warwick, Coventry, CV4 7AL, U.K.}

\author{S.~M.~Hamilton}
\author{W.~S.~Hopkins}
\author{S.~R.~Mackenzie }
\affiliation{Physical and Theoretical Chemistry Laboratory, South Parks Road, Oxford, OX1 3QZ, U.K.}

\author{P.~Gruene}
\affiliation{Fritz-Haber-Institut der Max-Planck-Gesellschaft,
Faradayweg 4-6, D-14195 Berlin, Germany}
\author{M.~Haertelt}
\affiliation{Fritz-Haber-Institut der Max-Planck-Gesellschaft,
Faradayweg 4-6, D-14195 Berlin, Germany}
\author{G.~Meijer}
\affiliation{Fritz-Haber-Institut der Max-Planck-Gesellschaft,
Faradayweg 4-6, D-14195 Berlin, Germany}
\author{A.~Fielicke}
\email{fielicke@fhi-berlin.mpg.de}
\affiliation{Fritz-Haber-Institut
der Max-Planck-Gesellschaft, Faradayweg 4-6, D-14195 Berlin,
Germany}

%

\date{\today}

\begin{abstract}
The geometric structure of the Rh$_8^+$ cation is investigated using a
combination of far-infrared multiple photon dissociation spectroscopy
 and density functional theory (DFT) calculations.
  The energetic ordering of the different structural motifs
is found to depend sensitively
on the choice of pure or hybrid exchange functionals. Comparison of
experimental and calculated spectra suggests the cluster to have a close-packed,
bicapped octahedral structure, in contrast to recent predictions of a cubic
structure for the neutral cluster. Our findings demonstrate the importance 
of including some exact exchange contributions in the DFT calculations, 
{\it via} hybrid functionals, when applied to rhodium clusters, and cast doubt 
on the application of pure functionals for late transition metal clusters 
in general.

\end{abstract}

\maketitle

The study of transition metal (TM) clusters offers the opportunity to probe the
fundamental physics involved in the transition from atomic to bulk
properties \cite{RevModPhys.77.371} and a means to better understand 
 potentially tractable model systems for supported catalysts\cite{knick}.  
A wide range of cluster properties, including
reactivity\cite{caeu05,rhN2O}, magnetic
moments\cite{clab94,payne:193401}, electric
polarizability\cite{beyer:104301} and ionization
potential\cite{mbksy90b,mbk03} have been found to depend
sensitively, and non-monotonically, on cluster size. These size-effects
reflect the complex evolution of the electronic and geometric structures 
and yet, the structures of most of these clusters are not currently 
known.

Although  knowledge of cluster structures is vital for developing a
deeper understanding of the observed properties, experimental
structure determination in the gas phase remains difficult.
Photoelectron spectroscopy has been applied to anionic
clusters and has yielded detailed information on the size-dependent evolution
of electronic structures\cite{wdw96}. Only in the last few 
years, however, has it become possible to probe more directly the
geometric structures of transition metal clusters in the gas phase,
e.g. by measuring their ion mobilities\cite{gwfak02}, by trapped ion electron
diffraction\cite{sbpihm05}, or far-infrared multiple photon dissociation
spectroscopy (FIR-MPD)\cite{AFGM_04_prl}.

Theory has been widely used to predict the geometric and electronic
structures  of clusters. Electronic structure theory is a vital aid 
to experimental efforts to determine cluster
structure\cite{AFGM_04_prl,PhilippGruene08012008,gwfak02,sbpihm05}
and to help understand the effects of structure on cluster 
properties\cite{rh6,GronbeckH._jp071117d}.
 The large
size of TM clusters, both in terms of the number of electrons
which must be treated and the complexity of the potential energy
surfaces (PES), generally restricts practical calculations to the
realm of DFT.  
 Application of DFT to mid-sized TM clusters requires caution
in the interpretation of results; not only because of the
approximations inherent to all contemporary exchange-correlation
functionals, but also because there is a lack of benchmark data from
high-quality multi-reference electronic structure theory, although
data does exist for rhodium clusters as large as the 
pentamer\cite{majumdar:2495}.
 Most
contemporary functionals were not designed with TM clusters in
mind and it is not clear if one single exchange-correlation functional is
to be preferred for all TM clusters.

Rhodium clusters present many of these challenges and opportunities 
and, as a consequence, have been studied experimentally in some 
detail\cite{clab94,caeu05,rhN2O,beyer:104301}. This work represents 
the first direct measurements of their structure.
Recent DFT calculations have predicted a range of unusual putative global
minimum and low-energy structures for rhodium 
clusters\cite{zxhpw04,bae,bae:125427,rh6,sun:043202}, which are
not based on the close-packed motifs identified for many TM
clusters\cite{fielicke05,frhm07,PhilippGruene08012008}. These
include a trigonal prism for Rh$_6$\cite{bae} and Rh$_6^+$\cite{rh6}
and  a cubic structure for Rh$_8$\cite{zxhpw04,bae,bae:125427}. 
Cubic motifs have 
also been reported for larger clusters\cite{bae:125427,sun:043202}. 
Cube-based  
structures have been rationalized by strong $d$-orbital character in
the bonding, favoring 90$^{\circ}$ bond angles and eight-center
bonding\cite{bae:125427}.
These structures are qualitatively different to the 
close-packed, polytetrahedral structures reported for most cluster 
sizes by  Futschek \etal\cite{0953-8984-17-38-001}.
 As the computational methods used in these studies are
rather similar, all using DFT,
 it is not clear if the differences in the results
are due to the range of structures and motifs
 considered for each cluster size or due to differences in the details of the
calculations, such as the exchange-correlation functionals used.

Cubic structures have also been predicted for other platinum
group metals including ruthenium \cite{wzhzlw04,li:045410}. For this
metal, Wang and Johnson\cite{WangL.-L._jp0555347} have investigated
in detail the effects of using pure and hybrid functionals on the
favored geometry of small ($n \le 4$) clusters. They reported that
while pure functionals predicted a square planar isomer as global
minimum for Ru$_4$, hybrid functionals favored a tetrahedral
isomer. They attributed the difference to changes in the relative
energy of the {\it s-} and {\it d-}orbitals in the ruthenium atom.
The proximity of rhodium and ruthenium in the periodic table suggested
similar effects may be observed for rhodium clusters, supported by the
recent results of Sun \etal \cite{sun:043202}.
Experiments which allow a direct
 comparison between the experimental results and calculated properties
therefore provide a vital means to  test and benchmark the
theoretical methods used to study these systems. Here, we focus on
 structure determination for the archetypal cubic
cluster, Rh$_8^+$, by gas-phase vibrational spectroscopy. An
eight-atom cluster is the smallest cluster which can form a complete
cube and therefore provides an ideal candidate system to test the
competition between cubic and close-packed geometries. IR
spectroscopy is particularly sensitive to symmetry through its
selection rules which, for an 8-atom cube, predict a single triply
degenerate IR-active mode of t$_{1u}$ symmetry.

FIR-MPD 
allows the measurement of size-specific IR spectra  of clusters in the
gas phase, which can be compared to the results of calculations. The
technique has been described in detail previously\cite{AFGM_04_prl}.
As implemented here, argon-tagged rhodium cluster cations are formed by laser
ablation of a rotating rhodium metal target. The resulting plasma
is cooled by a mixed pulse of helium and argon ({\it ca.} 0.3\%)
and carried down a cryogenically-cooled clustering channel (173 K)
before expansion into vacuum. The resulting cluster beam passes
through a skimmer before entering the FIR laser interaction region,
where it is intersected by a counterpropergating beam from 
the Free Electron Laser for Infrared eXperiments (FELIX).
By recording time-of-flight mass spectra alternately with and
without FELIX irradiation, the depletion of Ar-tagged clusters can
be measured as a function of IR wavelength. From this, the
size-specific absorption spectra are obtained as described
before\cite{AFGM_04_prl}.

\begin{figure}[ht]
\begin{center}
\includegraphics[width=8.5cm]{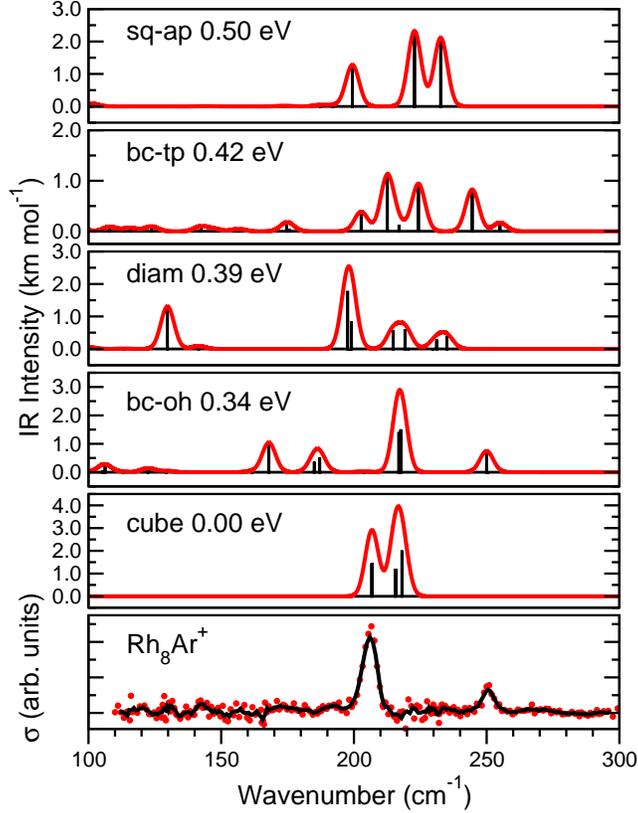}
\caption{\label{rh8spec-pure} Predicted IR spectra of Rh$_8^+$
from pure (PBE) DFT calculations
compared to the experimental FIR-MPD spectrum of Rh$_8$Ar$^+$, 
 $\sigma$ is the experimental IR cross section.
The labels correspond to the structures shown in Figure  \ref{fig:rh8pics}.}
\end{center}
\end{figure}

The experimental FIR-MPD spectrum of Rh$_8$Ar$^+$ is shown in the
lower panel of Figure \ref{rh8spec-pure}.  It appears relatively
simple, having an intense peak at  206$\pm$1 \wavno \ and a smaller
peak at  250$\pm$1 \wavno. The spectrum of Rh$_8$Ar$_2^+$ is very
similar, showing no large changes in peak position or intensity with
the degree of argon coverage. This indicates that the Ar acts as
spectator and is not influencing the cluster structures. This is
consistent with the findings, e.g. for the larger cobalt 
clusters\cite{gehrke:034306}.

In order to determine the structure of the clusters, the experimental
FIR-MPD spectra are compared with calculated vibrational spectra.
A thorough search of the PES has been performed in order to identify
the important, i.e. low-energy, geometric structures of the cluster.
We have used a two-stage approach;
first employing basin-hopping (BH) simulations\cite{dowa} to locate
candidate structures, and second by refining these candidate
structures. In the first stage, we used BH in two different
implementations; the first using the Sutton-Chen\cite{sutchen} model
potential, and the second based on the PES described by the local
density approximation\cite{PhysRev.81.385,vwn80}. 
In addition to the structures found in our BH simulations, a
range of previously reported structures were also
investigated\cite{bae,bae:125427,0953-8984-17-38-001}.
In the second stage all the 
 candidate structures were further optimized using DFT, performed
with the Gaussian03 package\cite{g03}. Both the pure PBE \cite{pbe}
and the hybrid PBE1 (25\% Hartree-Fock exchange)
\cite{pbe,perdew:9982} exchange-correlation functionals were used,
with the lanl2dz relativistic effective core potentials and
basis sets\cite{hay:270}. A range of spin-multiplicities were
considered in order to determine the favored spin-state for each
isomer and functional. The cluster symmetry was not constrained. All
the optimized structures featured C$_1$ symmetry, showing small
distortions from their higher-symmetry counterparts. 
Analytic frequency calculations
were performed to ensure the structures to be true local minima at
each level of theory  and to provide comparison with the FIR-MPD
spectra. To aid comparison, the calculated stick spectra were
 broadened by a Gaussian line shape function with a full width at half 
maximum height of 6
cm$^{-1}$, no scaling of the calculated frequencies was applied. 
The calculations were performed without explicit
consideration of the argon tagging atoms as the experimental spectra
were not found to depend significantly on the degree of argon
coverage.

\begin{table}
\caption{\label{tab:rh8en} Relative energies and favored spin-multiplicities
of different isomers of Rh$_8^+$ at pure and hybrid levels of theory.
** Structure collapsed during optimization.}
\centering
\begin{tabular}{|c|cc|cc|}
\hline
         &\multicolumn{2}{c|}{Pure} &\multicolumn{2}{c|}{Hybrid} \\
Isomer & 2$S$+1 & relative energy/eV  & 2$S$+1 & relative energy/eV \\
\hline
cube         & 8  & 0.00  &12   & 0.92  \\
bc-oh        &14  & 0.34  &14   & 0.00  \\
diam         &12  & 0.39  &**   & **  \\
bc-tp        &10  & 0.42  &14   & 0.18  \\
sq-ap        &10  & 0.50  &12   & 0.56  \\
\hline
\end{tabular}
\end{table}

\begin{figure}[ht]
\includegraphics[width=7.2cm]{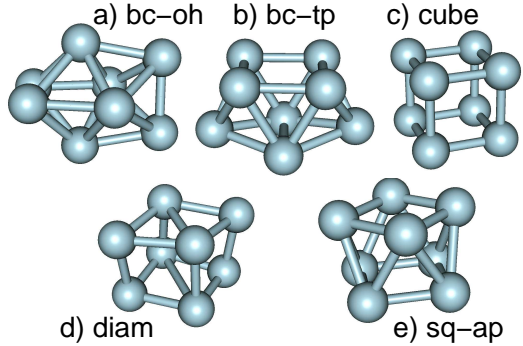}
\caption{ \label{fig:rh8pics} Geometric structures of low-energy
isomers of Rh$_8^+$ identified in DFT calculations: a) bicapped
octahedron (bc-oh), b) bicapped trigonal prism (bc-tp), c) cube, 
d) diamond prism (diam), e) square antiprism (sq-ap).}
\end{figure}

The results of the calculations are summarized in Figure
\ref{fig:rh8pics} which shows the most important geometric structures
and in Table \ref{tab:rh8en} where the relative energies of the
favored spin multiplicity of each isomer at the pure and hybrid
levels are listed. We find there to be a significant difference in
the energy ordering of the isomers at the two 
 levels. At the pure level, the cube is the lowest-energy
structure followed by the bicapped octahedron, broadly in agreement
with the results reported by Bae \etal for the neutral
clusters\cite{bae}.
At the hybrid level, however, the bicapped octahedron is favored, followed
by the bicapped trigonal prism. The cube isomer is supported at the
hybrid level, but is relatively high in energy (0.92 eV).
The favored spin-multiplicities for each isomer are generally
higher at the hybrid level, a point previously noted by Wang and
Johnson\cite{WangL.-L._jp0555347}. This is particularly evident for
the cube, for which the pure functional favors an octet while the
hybrid favors a 12-tet.  A large
number of unpaired electrons is generally in agreement with the
experimental finding of magnetic moments of up to 0.8$\pm$0.2 $\mu_B$
per atom for small neutral rhodium clusters\cite{clab94}. 
For ruthenium, significant differences
between pure and hybrid DFT emerge in the predicted
energetic ordering of the occupied orbitals of the atom
\cite{WangL.-L._jp0555347}. However, for rhodium we do not observe
such changes, making it more difficult to explain the underlying cause 
 of the differences in the pure and hybrid calculations.

\begin{figure}[ht]
\includegraphics[width=8.5cm]{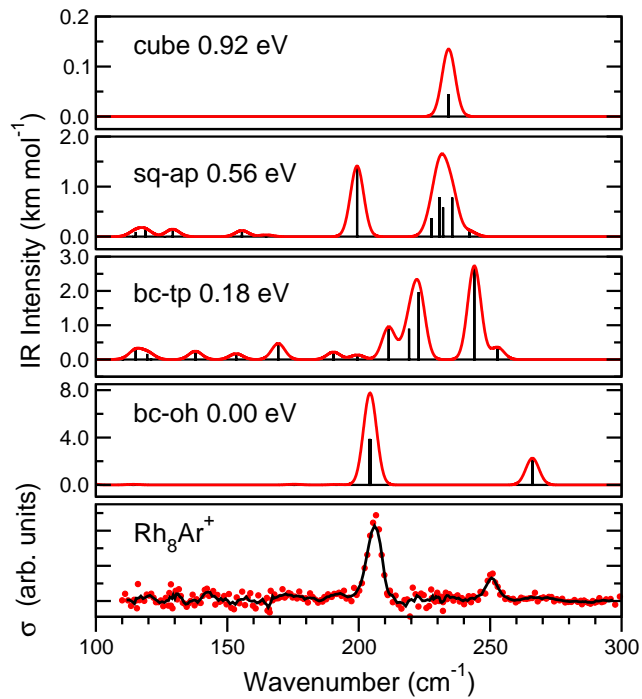}
\caption{\label{rh8spec-hyb} Predicted IR spectra of Rh$_8^+$
from  hybrid (PBE1) DFT calculations
compared to the experimental FIR-MPD spectrum of Rh$_8$Ar$^+$.
The labels correspond to the structures shown in Figure \ref{fig:rh8pics}.}
\end{figure}

In Figures \ref{rh8spec-pure} and \ref{rh8spec-hyb} we compare
the experimental FIR-MPD spectrum of Rh$_8$Ar$^+$ with the calculated
spectra from pure and hybrid calculations respectively. Comparison
of the experimental spectrum with calculated spectra at the pure
level show poor agreement with the spectrum of the cubic isomer.
For this, three rather closely spaced IR-active modes are predicted
at {\it ca.} 205 \wavno, but no features in the region of 250 \wavno.

A similar comparison with the spectra calculated at the hybrid level
shows the experimental spectrum to be well reproduced, both in
position and relative intensity, by the lowest-energy bicapped
octahedron isomer. This has an intense feature at 205 \wavno \ and a
weaker feature at 266 \wavno, blue-shifted by 15 \wavno \ compared to
the experimental spectrum.  It is notable that among the different
isomers at the pure level the best match to experiment is also
provided by the bicapped octahedron, though in this case 
extra features are predicted 
at low wavenumber. The calculated spectra of the other
isomers at the hybrid level do not provide a good match to
experiment, having too few or too many features. The spectrum of the
cubic isomer is calculated to have a triply degenerate feature of
very low oscillator strength at 234 \wavno, half way between the two
features observed experimentally.

The combination of observed vibrational spectra and calculations
based on the hybrid PBE1 functional strongly suggest that the Rh$_8^+$
cluster favors a close-packed, bicapped octahedron structure and not
a cubic structure as has been previously reported, and indicated by
our own DFT calculations based on the pure PBE functional. In the
absence of high-level multi-reference benchmark calculations, our
findings cast doubt over the suitability of pure functionals when
applied to small/mid-sized late TM clusters. Our evidence, while
specific to Rh$_8^+$, suggests in general that the expectation of open
structures such as those based on cubic motifs in a range of late TM
cluster sizes may be misconceived. Further evidence, comprising close
comparison of experiment and theory for other sizes of Rh clusters,
will be required to resolve this question.

{\bf Acknowledgements }
We gratefully acknowledge the support of the Warwick Centre for 
Scientific Computing for computer time 
and the Stichting voor Fundamenteel 
Onderzoek der Materie (FOM) for providing beam time on FELIX. The authors 
thank the FELIX staff for their skillful assistance, in particular Dr. 
B. Redlich and Dr. A.F.G. van der Meer. 
We thank Prof. David Rayner for the use his rhodium rod. 
 This work is supported by the 
Cluster of Excellence ``Unifying Concepts in
Catalysis'' coordinated by the Technische Universit\"{a}t Berlin and 
funded by the Deutsche Forschungsgemeinschaft.
DJH acknowledges an Early Career Fellowship from the Institute of Advanced 
Study, University of Warwick, SRM an Advanced Research Fellowship from the 
EPSRC and WSH the support of the Ramsay Memorial Fellowship Trust.

\end{document}